\documentclass[pdflatex,sn-mathphys-num]{sn-jnl}

\usepackage{graphicx}%
\usepackage{multirow}%
\usepackage{amsmath,amssymb,amsfonts}%
\usepackage{amsthm}%
\usepackage{mathrsfs}%
\usepackage[title]{appendix}%
\usepackage{xcolor}%
\usepackage{textcomp}%
\usepackage{manyfoot}%
\usepackage{booktabs}%
\usepackage{algorithm}%
\usepackage{algorithmicx}%
\usepackage{algpseudocode}%
\usepackage{listings}%
\usepackage{bm}
\usepackage{ulem}

\theoremstyle{thmstyleone}%
\theoremstyle{thmstyletwo}%

\theoremstyle{thmstylethree}%

\begin{document}

\title[Article Title]{Triplet correlations in superconductor/antiferromagnet heterostructures: dependence on type of antiferromagnetic ordering}


\author*[1]{\fnm{G.A.} \sur{Bobkov}}\email{gabobkov@mail.ru}

\author[1]{\fnm{V.A.} \sur{Bobkov}}\email{vyabobkov@mail.ru}

\author*[1,2]{\fnm{I.V.} \sur{Bobkova}}\email{ivbobkova@mail.ru}

\author[1]{\fnm{A.M.} \sur{Bobkov}}\email{ambobkov@mail.ru}

\author[1]{\fnm{A.A.} \sur{Golubov}}\email{a.golubov1960@gmail.com}

\affil*[1]{\orgname{Moscow Institute of Physics and Technology}, \orgaddress{\city{Dolgoprudny}, \postcode{141700}, \state{Moscow region}, \country{Russia}}}

\affil[2]{\orgname{National Research University Higher School of Economics}, \orgaddress{\city{Moscow}, \postcode{101000}, \country{Russia}}}


\abstract{
In recent years, a number of studies have predicted the emergence of a nontrivial proximity effect in superconductor/antiferromagnet (S/AF) heterostructures. This effect is of considerable interest for the efficient integration of antiferromagnetic materials into the fields of superconducting spintronics and electronics. A key element of this proximity effect is the Néel triplet correlations, initially predicted for S/AF heterostructures with checkerboard $G$-type antiferromagnetic ordering. However, various forms of antiferromagnetic ordering exist, and an important open question concerns the generalization of these results to such cases. In this paper, we develop a theory of the proximity effect in S/AF heterostructures with arbitrary two-sublattice antiferromagnetic ordering, aiming to clarify which antiferromagnets are capable of inducing triplet correlations and what structure these correlations may exhibit. We show that, in S/AF heterostructures with collinear compensated antiferromagnets, the dominant superconducting triplet correlations are of the checkerboard Néel type, as originally predicted for $G$-type antiferromagnets. In contrast, layered Néel triplet correlations—although potentially generated by layered antiferromagnets—are significantly weaker. Consequently, in S/AF heterostructures with layered antiferromagnetic ordering, the proximity-induced triplet correlations may exhibit either a checkerboard Néel or a conventional ferromagnetic structure, depending on the specific antiferromagnet and its orientation relative to the S/AF interface.}

\keywords{Superconducting proximity effect, Triplet correlations, Superconductor/antiferromagnet heterostructures, Superconducting spintronics}



\maketitle

\section{Introduction}

The superconducting proximity effect in mesoscopic heterostructures composed of conventional spin-singlet superconductors and ferromagnetic metals (S/F heterostructures) refers to the penetration of Cooper pairs from the superconductor into the adjacent ferromagnet, accompanied by a partial conversion of singlet pairs into spin-triplet pairs due to the presence of a macroscopic spin-splitting field in the ferromagnet \cite{Bergeret2005,Buzdin2005,Bergeret2018}. 

The typical length scale for the penetration of opposite-spin Cooper pairs into a ferromagnet is given by the magnetic coherence length $\xi_m = \hbar v_{f,F}/h$ for a ballistic ferromagnet and $\xi_m = \sqrt{\hbar D_F/h}$ for a diffusive one. This length typically ranges from a few nanometers to several tens of nanometers. In contrast, the penetration length of the equal-spin pairs is significantly longer and corresponds to the normal-state coherence length $\xi_N = \hbar v_{f,F}/2 \pi T$ for the ballistic ferromagnet and $\xi_N = \sqrt{\hbar D_F / 2 \pi T}$ in the diffusive regime \cite{Bergeret2005}. Simultaneously, triplet pairs are also induced within the superconductor due to the inverse proximity effect. These triplet pairs penetrate into the superconducting region over a distance comparable to the superconducting coherence length $\xi_S = \hbar v_{f,S}/2 \pi T_c$ for a ballistic superconductor and $\xi_S = \sqrt{\hbar D_S/2  \pi T_c}$ in the diffusive case. Here $v_{f,S(F)}$  denotes the Fermi velocity of electrons in the superconductor (ferromagnet), $D_{S(F)}$ is the diffusion coefficient, $h$ is the exchange field of the ferromagnet acting on the conduction electrons, $T$ is the temperature and $T_c$ is the critical temperature of the superconductor.

The same inverse proximity effect also arises at interfaces between a superconductor and a ferromagnetic insulator \cite{Eschrig2015}. The key distinction in this case is that, due to the insulating nature of the ferromagnet, the penetration of Cooper pairs into the ferromagnetic region is negligible. The formation of triplet pairs occurs at the expense of singlet pairs, thereby weakening the conventional superconducting state ~\cite{Chandrasekhar1962,Clogston1962,Sarma1963} for thin superconducting films with a thickness smaller than $\xi_S$ in contact with a ferromagnet, or results in a suppression of the superconducting order parameter (OP) near the interface over a distance on the order of $\xi_S$ in thicker films. The generation of triplet Cooper pairs in superconductor/ferromagnet (S/F) heterostructures has revealed a wealth of new and exciting physics \cite{Buzdin2005,Bergeret2005} and has provided a strong impetus for the development of superconducting spintronics \cite{Linder_review,Eschrig_review,Bergeret2018}, including the realization of superconducting spin valves \cite{DEGENNES1966,Li2013,Oh1997,Jara2014,Singh2015,Kamashev2019,Tagirov1999,Fominov2003,Zhu2010,Moraru2006,Singh2007,Banerjee2014,Leksin2012,Westerholt2005,Deutscher1969,Gu2002,Gu2015,Karminskaya2011,Wu2014,Halterman2015,Moen2017,Moen2018,Moen2020,Valls_book}, Josephson $\pi$-junctions \cite{Golubov2004,Buzdin2005} and $\varphi_0$-junctions \cite{Bobkova_review,Shukrinov_review,Kopasov2023}, superconducting spin currents carried by triplet Cooper pairs \cite{Grein2009,Brydon2013,Dahir2022,Ojajarvi2022,Jeon2020,Ouassou2019,Ouassou2017,Alidoust2010,Brydon2011,Gomperud2015,Jacobsen2016,Linder2017,Bobkova2017b,Bobkova2018,Aunsmo2024,Bobkov2024,Bobkov2025}, superconducting spin diode effect \cite{Schulz2025,Sun2025}, magnetization dynamics and magnetization switching induced by the triplet currents \cite{Bobkova2018,Zhu2004,Zhu2005,Holmqvist2011,Linder2011,Halterman2016,Kulagina2014,Linder2012,Takashima2017}, giant thermoelectric effects \cite{Machon2013,Ozaeta2014,Machon2014,Kolenda2016,Kolenda2017} and efficient low-temperature spin caloritronics \cite{Bergeret2018,Linder2016,Bobkova2017,Bobkova2021}, hybrid collective excitations \cite{Bobkova2022} and many other phenomena. 

On the other hand, antiferromagnets (AFs) are now attracting considerable attention as alternatives to ferromagnets for spintronics applications, owing to their robustness against external magnetic field perturbations, the absence of parasitic stray fields, and their ultrafast dynamics \cite{Baltz2018,Jungwirth2016,Brataas2020}. As discussed above, the majority of spintronics applications of S/F heterostructures rely on proximity-induced singlet-triplet conversion at the S/F interfaces. 

For a long time, it was widely accepted in the literature that S/AF heterostructures should behave similarly to S/N systems with respect to proximity-induced singlet–triplet conversion. This assumption is based on the fact that the net magnetization in an antiferromagnet, when averaged over the spatial extent of a typical Cooper pair, vanishes.
Accordingly, it was believed that only singlet pairs could penetrate into the antiferromagnetic metal, that triplet correlations could not be generated at a compensated S/AF interface with zero net magnetization, and that a superconductor interfaced with an antiferromagnetic metal or insulator via a compensated interface would not experience a net spin-splitting field or the associated suppression of the critical temperature~\cite{Werthammer1966}. In this framework, any macroscopic spin splitting in S/AF heterostructures was expected to arise only from an uncompensated interface magnetization \cite{Kamra2018}.

At the same time, it is important to note that in antiferromagnetic superconductors, the staggered exchange field suppresses superconductivity by opening a gap in the density of states \cite{Buzdin1986}. Another mechanism for superconductivity suppression in such systems is the behavior of nonmagnetic impurities, which act effectively as magnetic ones, thereby leading to the suppression of superconductivity \cite{Buzdin1986,Fyhn2022,Fyhn2022_1}, in full analogy with the effect of magnetic impurities in conventional singlet $s$-wave superconductors \cite{Abrikosov1961}.
Although these effects contribute to superconductivity suppression in antiferromagnetic superconductors and S/AF heterostructures, they are not associated with the physics of proximity-induced triplet superconducting correlations, which are the cornerstone of the phenomena and applications in S/F hybrids. Therefore, both from the perspective of fundamental physics and in view of potential applications in superconducting spintronics, the question of singlet–triplet conversion in S/AF heterostructures remains of significant interest and importance.

Recently it was found that in spite of the absence of a net magnetization, the checkerboard N\'eel magnetic order of the AF does induce spin-triplet correlations at S/AF interfaces, which penetrate into the superconductor and into the antiferromagnet (if it is metallic) \cite{Bobkov2022,Bobkova2024_neel_review} on the length scale of $\xi_S$ and $\xi_N$, respectively. Their amplitude flips sign from one lattice site to the next, just like the N\'eel magnetic order in the AF. These correlations were called N\'eel triplet Cooper pairs. The N\'eel triplet correlations exhibit physics that is in many respects similar to the physics of conventional triplet correlations in S/F heterostructures. In particular, they suppress superconductivity \cite{Bobkov2022}. The efficiency of the suppression is of the same order and even a bit stronger than the suppression by the conventional proximity-induced triplet correlations in S/F heterostructures. Also, the N\'eel triplet correlations provide an indirect exchange interaction in AF/S/AF trilayers \cite{Johnsen_Kamra_2023,Bobkov2024_spinvalve}, thus opening prospects for antiferromagnet-based superconducting spintronics. Similar to  S/F heterostructures \cite{Buzdin2005,Fominov2002,Radovic1991,Lazar2000,Vodopyanov2003,Jiang1995,Mercaldo1996,Muhge1996,Zdravkov2006,Zdravkov2010}, in S/AF heterostructures a mesoscopic FFLO state was predicted \cite{Bobkov2023_oscillations,Bobkov2024_singleimp,Bobkova2024_neel_review}, which experimentally should manifest itself as oscillations of the critical temperature of S/AF bilayers depending on the thickness of the AF layer \cite{Bobkov2023_oscillations}, which seem to be similar to existing experimental observations \cite{Bell2003,Hubener2002,Seeger2021,Wu2013}. However, the physical mechanism of the formation of the mesoscopic FFLO state in S/AF heterostructures \cite{Bobkova2024_neel_review} is not based on the nonzero macroscopic magnetization and is completely different from S/F heterostructures \cite{Demler1997}. By analogy with the S/F heterostructures \cite{Jacobsen2015,Ouassou2016,Simensen2018,Banerjee2018,Jonsen2019,Gonzalez-Ruano2020,Gonzalez-Ruano2021} the N\'eel triplets result in the anisotropy of the critical temperature of S/AF heterostructures in the presence of spin-orbit coupling (SOC) \cite{Bobkov2023_anisotropy}. 

At the same time, the N\'eel triplet correlations exhibit significant differences from conventional triplet correlations induced by the proximity effect in S/F heterostructures. Unlike in S/F heterostuctures, the amplitude of the N\'eel triplet correlations is highly sensitive to the value of the chemical potential (filling factor) in the material in which they are induced. Deviations from half filling suppress the amplitude of the N\'eel triplets for a given value of the antiferromagnet exchange field \cite{Bobkov2022}. In particular, if the antiferromagnetic exchange field $h$ is small with  respect to the chemical potential  $\mu$, the N\'eel triplets are of the order of $h/\mu \ll 1$. In with case they can be disregarded and any triplet correlations, both short-range and long-range, can only be of conventional type and are only produced if the S/AF interface is uncompemsated \cite{Fyhn2022_1,Salamone2024}. 

Depending on the value of the filling factor, the N\'eel triplets can be mainly interband (near half-filling) or intraband (far from half-filling) \cite{Bobkov2023_impurities}. This distinction leads to a qualitatively different  response of the N\'eel triplet correlations to nonmagnetic impurities: near half-filling they are suppressed by impurities, whereas far from half-filling they remain robust \cite{Bobkov2022,Bobkov2023_impurities,Bobkova2024_neel_review}. Moreover, all the physics of S/AF heterostructures is crucially sensitive to the value of the filling factor. In particular, the dependence of the critical temperature of S/AF heterostructures with canted AFs on the canting angle is opposite near half-filling and far from half-filling \cite{Chourasia2023}, the same applies to the magnetic anisotropy in S/AF heterostructures with SOC \cite{Bobkov2023_anisotropy}, to possibility to form Andreev bound states at single nonmagnetic impurities, which also depends on the filling factor \cite{Bobkov2024_singleimp}.

In this work, we demonstrate that the appearance of the N\'eel triplet correlations in antiferromagnetic superconductors and S/AF heterostructures is also highly sensitive to the type of the antiferromagnet order. All the physics discussed above was related to the checkerboard (G-type) antiferromagnetic order. In the present study, we extend the analysis to A-type and C-type (layered) antiferromagnets and compare the resulting superconducting proximity effects to those in S/F heterostructures and in S/AF heterostructures with checkerboard antiferromagnets. Our analysis includes both 2D and 3D geometries and considers different orientations of the antiferromagnet relative to the S/AF interface. Unexpectedly, we find that noticeable A-type or C-type N\'eel triplet correlations are not generated in antiferromagnetic superconductors and S/AF heterostructures even at S/AF interfaces with A-type or C-type antiferromagnets or in antiferromagnetic superconductors with A-type or C-type antiferromagnetic ordering and $s$-wave superconducting pairing. Instead, the layered antiferromagnets can only produce weak conventional triplet correlations or {\it checkerboard} triplet correlations depending on the orientation of the S/AF interface. Moreover, in some physical situations the proximity-induced triplets are not generated at all. Correspondingly, the suppression of the superconducting OP by such types of antiferromagnetism at S/AF interfaces is much weaker than in the S/F heterostructures or S/AF heterostructures with G-type antiferromagnets. We analyze the physical origin of these seemingly counterintuitive results.

The paper is organized as follows. In Sec.~\ref{model_formalism} we introduce the physical models under study and develop a generalized two-sublattice Green’s function formalism suitable for describing the superconducting proximity effect at interfaces with arbitrary two-sublattice antiferromagnets. In Sec.~\ref{triplets} we present our results on proximity-induced triplet correlations, including analytical results for homogeneous antiferromagnetic superconductors in Sec.\ref{homogeneous} and numerical results for 2D and 3D S/AF heterostructures in Sec.\ref{numerical}. In Sec.\ref{OP}, we analyze and compare the suppression of the superconducting OP by different types of antiferromagnetic order. Our conclusions are summarized in Sec.\ref{conclusions}.

\section{Model and generalized two-sublattice formalism}
\label{model_formalism}

For an arbitrary two-sublattice antiferromagnetic order the corresponding antiferromagnetic superconductors and S/AF heterostructures can be described by the following tight-binding Hamiltonian: 
\begin{align}
\hat H= - \sum \limits_{ \bm i\bm j\nu \nu' \alpha} t_{\bm i\bm j}^{\nu\nu'} \hat \psi_{\bm i \alpha}^{\nu\dagger} \hat \psi_{\bm j \alpha}^{ \nu'} + \sum \limits_{\bm i,\nu } (\Delta_{\bm i}^\nu \hat \psi_{\bm i\uparrow}^{\nu\dagger} \hat \psi_{\bm i\downarrow}^{\nu\dagger} + H.c.) + \sum \limits_{\bm i \nu,\alpha \beta} \hat \psi_{\bm i\alpha}^{\nu \dagger} (\bm h_{\bm i}^\nu \bm \sigma)_{\alpha \beta} \hat \psi_{\bm i\beta}^\nu,
\label{ham}
\end{align}
where $ \bm \sigma = (\sigma_x,\sigma_y,\sigma_z)^T$ is the vector of Pauli matrices in spin space, $\nu=A,B$ is the sublattice index, $\hat{\psi}_{\bm i \sigma}^{\nu \dagger}(\hat{\psi}_{\bm i \sigma}^{\nu })$ is the creation (annihilation) operator for an electron with spin $\sigma=\uparrow,\downarrow$ at the sublattice $\nu$ of the unit cell $\bm i$ what we will denote as $(\bm i,\nu)$. $t_{\bm i\bm j}^{\nu\nu'}$ parameterizes the hopping between sites $(\bm i,\nu)$ and $(\bm j,\nu')$, $\Delta_{\bm i}^\nu$ accounts for on-site s-wave pairing. For homogeneous antiferromagnetic superconductors it is nonzero at all sites of the system, and for S/AF heterostructures it is only nonzero in the S layer. It is worth mentioning that we only consider antiferromagnetic superconductors with $s$-wave pairing \cite{Buzdin1986} here. At the same time, other types of superconducting pairing can also be realized in antiferromagnetic superconductors. In particular, if one studies a spin-fluctuation-mediated superconductivity in the framework of the Habbard model, the resulting dominating pairing can have a $d$-wave momentum symmetry \cite{Romer2016}. In this case study of the singlet-triplet conversion is a separate problem, which is beyond the scope of our work. $\hat n_{\bm i \sigma}^\nu = \hat \psi_{\bm i \sigma}^{\nu\dagger} \hat \psi_{\bm i \sigma}^\nu$ is the particle number operator at the site $(\bm i,\nu)$. $h_{\bm i}$ is an exchange term induced by localized magnetization at the site $(\bm i, \nu)$. Again, for homogeneous antiferromagnetic superconductors it is nonzero at all sites of the system, and for S/AF heterostructures it only presents inside the antiferromagnet. Note that the filling factor, often denoted as $\mu_{\bm i}$, is also included in this first term (on-site term $\mu_{\bm i}=2t_{\bm i\bm i}^{\nu\nu}$).

The Matsubara Green's function in the two-sublattice formalism is $8 \times 8$ matrix in the direct product of spin, particle-hole and sublattice spaces. Introducing the two-sublattice Nambu spinor $\check \psi_{\bm i} = (\hat \psi_{\bm i,\uparrow}^A, \hat \psi_{\bm i,\downarrow}^A, \hat \psi_{\bm i,\uparrow}^B,\hat \psi_{\bm i,\downarrow}^B, \hat \psi_{\bm i,\uparrow}^{A\dagger}, \hat \psi_{\bm i,\downarrow}^{A\dagger}, \hat \psi_{\bm i,\uparrow}^{B\dagger}, \hat \psi_{\bm i,\downarrow}^{B\dagger})^T$ we define the Green's function as follows: 
\begin{eqnarray}
\check G_{\bm i \bm j}(\tau_1, \tau_2) = -\langle T_\tau \check \psi_{\bm i}(\tau_1) \check \psi_{\bm j}^\dagger(\tau_2) \rangle,
\label{Green_Gorkov}
\end{eqnarray}
where $\langle T_\tau ... \rangle$ means imaginary time-ordered thermal averaging. For the system described by Hamiltonian (\ref{ham}) the Heisenberg equation of motion for spinor $\check \psi_{\bm i}$ takes the form: 
\begin{eqnarray}
\frac{d\check \psi_{\bm i}}{d \tau} = [\hat H, \check \psi_{\bm i}] = 
\tau_z \Bigl( { \hat t} - \check \Delta_{\bm i} i\sigma_y -  \hat {\bm h_{i}} \check {\bm \sigma}   \Bigr)\check \psi_{\bm i},
\label{psi_motion}
\end{eqnarray}
where $\sigma_k$, $\tau_k$ and $\rho_k$ ($k=x,y,z$) are Pauli matrices in spin, particle-hole and sublattice spaces, respectively. $\hat {\bm h}=\bm h^A(1+\rho_z)/2+\bm h^B(1-\rho_z)/2$, $\check {\bm \sigma} = \bm\sigma (1+\tau_z)/2 + \bm\sigma^* (1-\tau_z)/2$ is the quasiparticle spin operator and $\check \Delta_{\bm i} = \Delta_{\bm i} \tau_+ + \Delta_{\bm i}^* \tau_-$ with $\tau_\pm  = (\tau_x \pm i \tau_y)/2$. Here we assume that $\Delta_{\bm i}^A = \Delta_{\bm i}^B = \Delta_{\bm i}$, that is the order parameter values are equal for $A$ and $B$ sites of a unit cell which is a consequence of 
equivalence of two sublattices and is confirmed by the results of the numerical solution of the self-consistency equation. The operator $\hat t$ is defined as
\begin{eqnarray}
\hat t \check \psi_{\bm i} = \sum \limits_{\bm j} \left(
\begin{array}{cc}
t_{\bm i \bm j}^{AA} & t_{\bm i \bm j}^{AB} \\
t_{\bm i \bm j}^{BA} & t_{\bm i \bm j}^{BB}
\end{array}
\right)_\rho \check \psi_{\bm j}.
\label{j}
\end{eqnarray}
The Green's function Eq.~(\ref{Green_Gorkov}) obeys the following equation:
\begin{eqnarray}
\frac{d \check G_{\bm i \bm j}}{d \tau_1} = -\delta (\tau_1-\tau_2)\delta_{\bm i \bm j} - \langle T_\tau \frac{d \check \psi_{\bm i}(\tau_1)}{d \tau_1} \check \psi_{\bm j}^\dagger(\tau_2)\rangle .
\label{Green_gorkov_eq}
\end{eqnarray}
Substituting Eq.~(\ref{psi_motion}) into Eq.~(\ref{Green_gorkov_eq}) and expressing the Green's function in terms of the Matsubara frequencies $\omega_m = \pi T(2m+1)$ we obtain:
\begin{align}
G_{\bm i}^{-1} \check G_{\bm i \bm j}(\omega_m) = \delta_{\bm i \bm j}, \label{gorkov_eq_ml} 
\end{align}
\begin{align}
G_{\bm i}^{-1} = \tau_z ( \hat t  - \check \Delta_{\bm i} i \sigma_y - \hat {\bm h}_{\bm i} \check {\bm \sigma} )+ i \omega_m .
\label{G_i}
\end{align}
Further we define the following transformed Green's function, which allows us to present the Gor'kov equation in a more convenient form:
\begin{align}
\check {\tilde G}_{\bm i\bm j} = 
\left(
\begin{array}{cc}
1 & 0 \\
0 & -i\sigma_y
\end{array}
\right)_\tau \check G_{\bm i\bm j} 
\left(
\begin{array}{cc}
1 & 0 \\
0 & -i\sigma_y
\end{array}
\right)_\tau ,
\label{unitary}
\end{align}
and resulting Gor'kov equation takes the form:
\begin{align}
\left[ i \omega_m \tau_z + \tau_z \check \Delta_{\bm i} - \hat {\bm h}_{\bm i} \bm \sigma  \tau_z \right] \check {\tilde G}_{\bm i \bm j} +
\sum_{\bm k}\left(
\begin{array}{cc}
t_{\bm i\bm k}^{AA} & t_{\bm i\bm k}^{AB} \\
t_{\bm i\bm k}^{BA} & t_{\bm i\bm k}^{BB}
\end{array}
\right)_\rho \check {\tilde G}_{\bm k \bm j}  = \delta_{\bm i\bm j} .
\label{gorkov} 
\end{align}
The superconducting order parameter is determined self-consistently according to the following equation:
\begin{align}
\Delta_{\bm i}= \gamma\langle\hat \psi_{\bm i \downarrow} \hat \psi_{\bm i \uparrow} \rangle =  \gamma \sum\limits_{|\omega_m|<\Omega_D} {\rm Tr}[ \check {\tilde G}_{\bm i\bm i} (\tau_x-i\tau_y)\rho_0]/4,
\label{eq:self_consist_BDG_green}
\end{align}
where $\gamma$ is the coupling constant and $\Omega_D$ is the Debye energy.

\section{Triplet correlations induced by proximity to two-sublattice antiferromagnets of different types}
\label{triplets}

\subsection{Triplet correlations in a homogeneous antiferromagnetic superconductor}
\label{homogeneous}

Here we consider a homogeneous antiferromagnetic superconductor. Since it is a fully translationally invariant problem, the Green's function depends only on the difference $\bm i-\bm j$. The Fourier transform on this difference takes the form:
\begin{align}
\check {\tilde G}(\bm p) = \mathbb{F}(\check {\tilde G}_{\bm i \bm j}) = \int d^3 (\bm i-\bm j)  e^{-i \bm p(\bm i - \bm j)}e^{\frac{-\displaystyle ip_x a_x \rho_z}{\displaystyle 2}}\check {\tilde G}_{\bm i \bm j} e^{\frac{\displaystyle ip_x a_x \rho_z}{\displaystyle 2}}.
\label{mixed}
\end{align}
We assume that the unit cell containing two sites belonging to $A$ and $B$ sublattices is chosen along the $x$-axis (and it is the origin of additional exponents ${\rm exp}(\displaystyle \pm ip_x a_x \rho_z/2)$). For the momentum-dependent Green's function $\check {\tilde G}(\bm p)$ Eq.~(\ref{gorkov}) takes the form:
\begin{align}
\left[ i \omega_m \tau_z + \tau_z \check \Delta - \hat {\bm h}  \bm \sigma  \tau_z \right] \check {\tilde G}(\bm p) + \left(
\begin{array}{cc}
\xi^{AA}(\bm p) & \xi^{AB}(\bm p) \\
\xi^{BA}(\bm p) & \xi^{BB}(\bm p)
\end{array}
\right)_\rho \check {\tilde G}(\bm p)  = 1 .
\label{gorkov_final} 
\end{align}
where $\xi^{\nu\nu'}(\bm p)=\sum_{\bm j} t_{\bm j\bm 0}^{\nu\nu'}e^{i\bm p\bm j}$. In this paper we study compensated collinear antiferromagnets with equivalent sublattices and do not take into account such features as spin-orbit interaction, so we can write: $\hat {\bm h}=\bm h \rho_z$, $\xi^{AA}=\xi^{BB}=\xi_D$, $\xi^{AB}=\xi^{BA}=\xi_O$. For clarity, let us consider specific cases: 

{\it Example 1}: 3D cubic (with lattice constant $a$) checkerboard  antiferromagnet (G-type) on a square lattice with only nonzero nearest-neighbor hopping element $t$ and on-site energy $\mu$. In this case $\xi_D(\bm p)=\mu$, $\xi_O(\bm p)=2t[\cos(p_x a)+\cos(p_y a)+\cos(p_z a)]$.  

{\it Example 2}: 3D cubic layered antiferromagnet with ferromagnetic coupling in the plane of the layers and antiferromagnetic interlayer coupling (A-type, ferromagnetic layers are in the ${\rm yz}$-plane) on a square lattice with only nonzero nearest-neighbor hopping element $t$. Then $\xi_D(\bm p)=2t[\cos(p_y a)+\cos(p_z a)]+\mu$, $\xi_O(\bm p)=2t\cos(p_x a)$.

{\it Example 3}: 3D cubic antiferromagnet with ferromagnetic coupling along the $z$-axis and antiferromagnetic coupling in the $xy$ plane (C-type antiferromagnet) on a square lattice with only nonzero nearest-neighbor hopping element $t$. Then $\xi_D(\bm p)=2t\cos(p_z a)+\mu$, $\xi_O(\bm p)=2t[\cos(p_x a)+\cos(p_y a)]$.

Eq.~(\ref{gorkov_final}) can be easily solved. All components of the anomalous Green's function $F_\alpha={\rm Tr}[\check {\tilde G} (\tau_x-i\tau_y) \rho_\alpha]/2$ in the sublattice space take the form:
\begin{align}
    F_0=\frac{-\Delta (-2h^2+2\Delta^2+\xi_1^2+\xi_2^2+2\omega_m^2)/2}{h^4-2h^2(\Delta^2+\xi_1\xi_2-\omega_m^2)+(\Delta^2+\xi_1^2+\omega_m^2)(\Delta^2+\xi_2^2+\omega_m^2)}  \nonumber \\
    F_x=\frac{\Delta (\xi_1-\xi_2)(\xi_1+\xi_2)/2}{h^4-2h^2(\Delta^2+\xi_1\xi_2-\omega_m^2)+(\Delta^2+\xi_1^2+\omega_m^2)(\Delta^2+\xi_2^2+\omega_m^2)}  \nonumber \\
    F_y=\frac{i\bm h\bm \sigma (\xi_1-\xi_2)\Delta}{h^4-2h^2(\Delta^2+\xi_1\xi_2-\omega_m^2)+(\Delta^2+\xi_1^2+\omega_m^2)(\Delta^2+\xi_2^2+\omega_m^2)}  \nonumber \\
    F_z=\frac{2i\bm h\bm \sigma \omega_m\Delta}{h^4-2h^2(\Delta^2+\xi_1\xi_2-\omega_m^2)+(\Delta^2+\xi_1^2+\omega_m^2)(\Delta^2+\xi_2^2+\omega_m^2)} ,
    \label{gorkov_solution}
\end{align}
where $\xi_{1,2}=(\xi_D\pm\xi_O)$. The dispersion relation of electrons in the superconducting antiferromagnet can be taken from the pole of the Green's function Eq.~(\ref{gorkov_solution}). In the normal state $\Delta = 0$ the electron dispersion takes the form $\varepsilon=-\frac{1}{2}(\xi_D\pm \sqrt{\xi_O^2+h^2})$. Thus, physically $-\xi_{1,2}$ are two branches of the electron dispersion at $h=0$. 


The main contribution to any observables related to superconductivity comes from the region in the momentum space in the vicinity of the Fermi surface. Since in most part of experimentally relevant situations $(h, \Delta, T) \ll t$, the region of interest in the momentum space is determined by the condition $\xi_1 \ll t$ or $\xi_2 \ll t$. Then for an arbitrary compensated collinear antiferromagnetic superconductor there are three options: 

\begin{figure}[t]
\centering
\includegraphics[width=0.8\columnwidth]{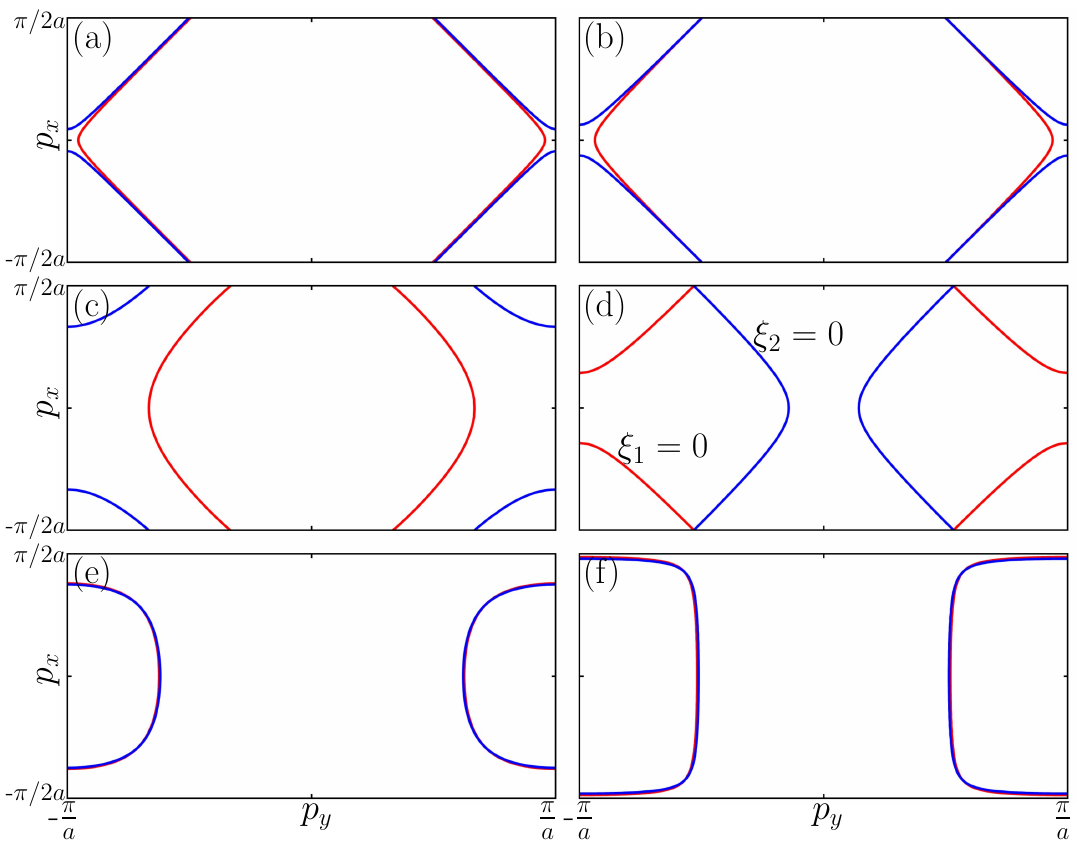}
\caption{2D Fermi surfaces for normal state of antiferromagnetic superconductors with different types of magnetic ordering. $h$ is assumed to be small and disregarded in the figures. A square crystal lattice is assumed. For all panels the Fermi surface sheets originating from $\xi_{1(2)}=0$ are shown in red (blue) color. (a) G-type AF order, $\mu=-0.02t$, the next-nearest-neighbor hopping element $t_2 = 0$. (b) G-type AF order, $t_2=0.01t$, $\mu=0$. (c) G-type AF order, $\mu=-t$, $t_2=0$. (d) A-type AF order, $\mu=0.2t$, $t_2 = 0$. (e) G-type AF order, $t=0.02t_2$, $\mu=0.2t_2$. (f) G-type AF order, $t=0.02t_2$, $\mu=1.5t_2$.} 
 \label{fig:examples}
\end{figure}

I) $\xi_O\gg\xi_D$. This case corresponds to {\it Example  1} of the checkerboard antiferromagnet at $\mu \ll t$, that is near half-filling. In this case, in the vicinity of the Fermi surface two Fermi surfaces corresponding to $\xi_{1,2} = 0$ nearly coincide, which means that there is a good nesting between them.   In the vicinity of the Fermi surface one can write $\xi_1=\xi+\delta \xi$, $\xi_2=-\xi+\delta \xi$, where $\delta\xi$ is small, but possibly non-zero due to $\xi_D\neq 0$. The examples of the corresponding Fermi surfaces are presented in Figs.~\ref{fig:examples}(a) and (b). The deviation from the ideal nesting, that is $\xi_D \neq 0$ in  Fig.~\ref{fig:examples}(a) arises due to non-zero $\mu$, and in Figs.~\ref{fig:examples}(b) due to the small nonzero value of the next-nearest-neighbor hopping element. At $\xi_D = 0$:
\begin{align}
    F_x=0~~~~~~~~~~~~~~~~~~~~~~~~~~~~ \nonumber\\ 
    F_0=\frac{-\Delta (-h^2+\Delta^2+\xi^2+\omega_m^2)}{((h-\Delta)^2+\xi^2+\omega_m^2)((h+\Delta)^2+\xi^2+\omega_m^2)} \nonumber\\ 
    F_y=\frac{2i\Delta \bm h\bm \sigma \xi}{((h-\Delta)^2+\xi^2+\omega_m^2)((h+\Delta)^2+\xi^2+\omega_m^2)} \nonumber\\ 
    F_z=\frac{2i\Delta \bm h\bm \sigma \omega_m}{((h-\Delta)^2+\xi^2+\omega_m^2)((h+\Delta)^2+\xi^2+\omega_m^2)}
    \label{triplets_G}
\end{align}
The on-site N\'eel triplet correlations are described by $F_z$. From Eq.~(\ref{triplets_G}) it is seen that the conventional singlet correlations and the N\'eel triplet correlations are of the same order of magnitude in the vicinity of the Fermi surface. Also the exchange field $h$ enters the singlet component $F_0$ suppressing it, what is the result of the singlet-triplet conversion. 

II) $\xi_O\sim\xi_D$. This case corresponds to {\it Examples  2,3} of the $A$- and $C$-type antiferromagnets or a $G$-type antiferromagnet far from half-filling. In this case, in the vicinity of some parts of the Fermi surface $\xi_1 \ll t$ and $\xi_2 \sim t$. In other parts of the Fermi surface we have the opposite situation $\xi_2 \ll t$ and $\xi_1 \sim t$. Please note that $t$ is the characteristic energy width of the conduction band. In other words, the Fermi surface sheets are far from each other and the system is far from the nesting condition. The examples of the corresponding Fermi surfaces are presented in Figs.~\ref{fig:examples}(c)-(d). In Fig.~\ref{fig:examples}(c) the case of the $G$-type antiferromagnet far from half-filling is demonstrated, and Fig.~\ref{fig:examples}(d) corresponds to the case of $A$-type antiferromagnet. In the case $\xi_1 \ll t$ and $\xi_2 \sim t$ one can write
\begin{align}
    F_0=-\frac{\Delta/2}{\Delta^2+\xi_1^2+\omega_m^2}+&{\rm O}[\frac{1}{t}], \nonumber\\ 
     F_{x,y,z}={\rm O}[\frac{1}{t}]&,
     \label{striped_1}
\end{align}
and for $\xi_2 \ll t$ and $\xi_1 \sim t$
\begin{align}
    F_0=-\frac{\Delta/2}{\Delta^2+\xi_1^2+\omega_m^2}+&{\rm O}[\frac{1}{t}],  \nonumber \\ 
     F_{x,y,z}={\rm O}[\frac{1}{t}]&.
    \label{striped_2}
\end{align}
Up to the leading order with respect to the parameter ${\rm max}(h, \Delta, T)/t$ solutions for the anomalous Green's functions expressed by Eqs.~(\ref{striped_1}) and (\ref{striped_2}) coincide with the solution for a conventional superconductor in the absence of an exchange field. This means that in this case the triplet correlations of {\it any} type are not produced to the considered accuracy. For the majority of materials and especially for thin-film S/AF heterostructures, which are frequently considered in the framework of the simplified model of the homogeneous antiferromagnetic superconductor \cite{Bobkov2022,Bobkov2023_impurities,Sukhachov2025} the condition ${\rm max}(h, \Delta, T)/t \ll 1$ is fulfilled with good accuracy because $t$ is of the order of eV, while $h$ varies from $1$ up to $100$ meV and $\Delta$ is typically even smaller. Since the triplet correlations are not produced and the singlet-triplet conversion is absent, there is $\it no$ suppression of superconductivity. More precisely, it is small in parameter $[h/t]^2$. 

It is important to note that there is no any fundamental symmetry reasons, which prohibit the existence of triplet correlations in the considered case $\xi_D \sim \xi_O$. Indeed, we have numerically check that the checkerboard N\'eel triplet correlations are generated in the case of the $G$-type antiferromagnetic order and $\mu \sim t$, but their amplitude is of the first order with respect to the small parameter $h/\mu \sim h/t$. Analogously, for the case of $A$-type and $C$-type antiferromagnets the corresponding $A$-type or $C$-type N\'eel triplet correlations appear, but their amplitude is again controlled by the same parameter. Please note that unlike the case of $G$-type antiferromagnet for the $A$-type or $C$-type antiferromagnet the implementation of the nesting condition cannot be achieved under any parameters. For this reason at $h \ll t$ we cannot not observe $A$-type or $C$-type N\'eel triplet correlations in contrast to the checkerboard N\'eel triplet superconducting correlations.

III) $\xi_O\ll\xi_D$. It is quite a specific case from the point of view of real materials. It can be realized, for example, in the case of checkerboard antiferromagnetic order and if the nearest-neighbor hopping element is negligible with respect to the next-nearest hopping element. In this case two sublattices of the antiferromagnetic material are practically not connected to each other.  The corresponding Fermi surfaces are shown in Figs.~\ref{fig:examples}(e) and (f). Both panels are plotted for the same dominating value of the next-nearest hopping element, but for different values on the on-site energy $\mu$. By comparing  Figs.~\ref{fig:examples}(e) and (f) one can conclude that in this case $\mu$ does not have a qualitative effect on the Fermi surfaces. Blue sheets originate from one of the sublattices and red sheets are produced by the other. They coincide because both sublattices are equivalent and do not interact. Accordingly, in each of them the situation is equivalent to the proximity effect in a ferromagnetic superconductor or a superconductor in a Zeeman field \cite{Sarma1963}.  

\subsection{Numerical results for triplet correlations in a S/AF heterostructure}
\label{numerical}

\subsubsection{BdG approach}
\label{BdG}
In this section we study the triplet correlations induced at S/AF interfaces, that is in the case when superconductivity and antiferromagnetism are spatially separated. Namely, we consider a S/AF heterostructure infinite along the $y$ and $z$ axes for the 3D case or along the $y$ axis for the 2D case in the simplest tight-binding model on a square lattice with only nonzero nearest-neighbor hopping element $t$. Also we restrict ourselves by the half-filling case $\mu=0$ in order to avoid Friedel oscillations of the correlation functions and absolutely transparent S/AF interface. This spatially inhomogeneous problem is studied numerically and for the numerical calculations we do not use the two-sublattice formalism described above. Instead, we choose a box of 2 atoms along the $y$-axis 
direction in 2D (of $2 \times 2$ atoms in the $yz$-plane for 3D), then make a Fourier transform along the S/AF interface and consider the full coordinate dependence along the $x$-axis. This approach is fully equivalent to the two-sublattice formalism, but a bit more convenient for numerical calculations. Then the corresponding Hamiltonian takes the form:
\begin{align}
\hat H= - t \sum \limits_{ \langle \bm i \bm j \rangle \alpha} \hat \psi_{\bm i \alpha}^{\dagger} \hat \psi_{\bm j \alpha} + \sum \limits_{\bm i } (\Delta_{\bm i} \hat \psi_{\bm i\uparrow}^{\dagger} \hat \psi_{\bm i\downarrow}^{\dagger} + H.c.) + \sum \limits_{\bm i,\alpha \beta} \hat \psi_{\bm i\alpha}^{ \dagger} (\bm h_{\bm i} \bm \sigma)_{\alpha \beta} \hat \psi_{\bm i\beta},
\label{eq_hamBDG}
\end{align}
where $\langle \bm i \bm j \rangle $ means summation over the nearest neighbors. We diagonalize Hamiltonian (\ref{eq_hamBDG}) by the Bogoliubov transformation:
\begin{align}
\psi_{\bm i\sigma}=\sum\limits_n u_{n\sigma}^{\bm i}\hat b_n+v^{\bm i *}_{n\sigma}\hat b_n^\dagger, 
\label{bogolubov}
\end{align}
Then the resulting Bogoliubov – de Gennes equations take the form:
\begin{align}
 \sigma \Delta_{\bm i} v^{ \bm i}_{n,-\sigma} - t \sum\limits_{\langle\bm i \bm j\rangle}  u^{\bm j}_{ n, \sigma} + \sum_{\alpha}(\bm h_{\bm i} \bm \sigma)_{\alpha\sigma} u^{\bm i}_{ n, \alpha}  & = \varepsilon_n u_{n,\sigma}^{\bm i} \nonumber \\  
\sigma \Delta_{ \bm i}^* u^{\bm i}_{n,-\sigma} - t \sum\limits_{\langle\bm i \bm j\rangle}  v^{\bm j}_{ n, \sigma} - \sum_{\alpha}(\bm h_{\bm i} \bm \sigma)_{\alpha\sigma} u^{\bm i}_{ n, \alpha} & = -\varepsilon_n v_{n,\sigma}^{\bm i}, 
\label{bdg}
\end{align}
The superconducting order parameter in Eqs.~(\ref{bdg}) is calculated self-consistently:
\begin{align}
\Delta_{\bm i}= \gamma\langle\hat \psi_{\bm i \downarrow} \hat \psi_{\bm i \uparrow} \rangle =  \gamma \sum\limits_{|\varepsilon_n|<\Omega_D} (u_{n,\downarrow}^{\bm i} v_{n,\uparrow}^{\bm i*}(1-f_n)+u_{n,\uparrow}^{\bm i} v_{n,\downarrow}^{\bm i*}f_n),
\label{eq:self_consist_BDG}
\end{align}
where $\gamma$ is the coupling constant and $\Omega_D$ is the Debye energy. The quasiparticle distribution function is assumed to be the equilibrium Fermi distribution $f_n = \langle b_n^\dagger b_n \rangle = 1/(1+e^{\varepsilon_n/T})$.

To investigate the structure of superconducting  correlations one need to calculate the anomalous Green's function, which contains information about both singlet and triplet superconducting correlations occurring at the interface. The anomalous Green's function in Matsubara representation can be calculated as $F_{\bm i, \alpha \beta} = - \langle T_\tau \hat \psi_{\bm i \alpha}(\tau) \hat \psi_{\bm i \beta}(0) \rangle$, where $\tau$ is the imaginary time. The component of this anomalous Green's function for a given Matsubara frequency $\omega_m = \pi T(2m+1)$ is calculated as follows: 
\begin{align}
F_{\bm i,\alpha\beta}(\omega_m)= \sum\limits_n (\frac{  u_{n,\alpha}^{\bm i} v_{n,\beta}^{\bm i*}}{i \omega_m -\varepsilon_n}+\frac{ u_{n,\beta}^{\bm i} v_{n,\alpha}^{\bm i*}}{i \omega_m +\varepsilon_n}).
\label{eq:anom_func_BDG}
\end{align}
Only off-diagonal in spin space components, corresponding to opposite-spin pairs, are nonzero for the considered case of collinear antifferomagnet. The singlet (triplet) correlations are described by $F_{\bm i}^{s,t}(\omega_m) = F_{\bm i,\uparrow \downarrow}(\omega_m) \mp F_{\bm i,\downarrow \uparrow}(\omega_m)$. It is important to note that on-site triplet correlations are odd in Matsubara frequency, as it should be according to the general fermionic symmetry. Therefore we calculate the sum over only the positive Matsubara frequencies 
\begin{align}
F_{\bm i}^{s(t)} = \sum \limits_{0<\omega_m<\Omega_D} F_{\bm i}^{s(t)}(\omega_m).
\label{triplet_total}
\end{align}
Since the problem is translationally invariant, we choose a box of 2 atoms in the $y$ direction in 2D (of $2\times 2$ atoms in the $yz$ plane for 3D) and also consider the full coordinate dependence along the $x$-axis. The spatial dependence of the solution along the $y$ and $z$ directions takes the form:
\begin{align}
    u(v)_{n,\sigma}^{\bm i}(\bm p)=u(v)_{n,\sigma}^{\bm \eta} e^{i (p_y i_y+ p_z i_z)},
\end{align}
where $\bm \eta$ is the projection of $\bm i$ on the chosen translationally invariant box. 
In this case, solution of Eq.~(\ref{bdg}) is reduced to a problem of finding the eigenvalues and eigenvectors of $2^{D+1}(N_S+N_{AF})\times 2^{D+1}(N_S+N_{AF})$ matrix (here $D$ is the dimensionality of the problem and $N_S,N_{AF}$ is the number of sites along $x$-axis for superconducting and magnetic parts of the system). The eigenvalues of this matrix are the energies $\varepsilon_n$ and its eigenvectors consist of  the amplitudes $u_{n,\sigma}^{\bm \eta},v_{n,\sigma}^{\bm \eta}$ for all sites $\bm \eta$. By numerically diagonalizing the matrix, we find its eigenenergies and eigenvectors. Then Eqs.~(\ref{eq:self_consist_BDG}) and (\ref{eq:anom_func_BDG}) take the form:
\begin{align}
\Delta_{\bm \eta}= \gamma \int\limits_{BZ} \frac{d(pa)^{D-1}}{\pi^{D-1}}\sum\limits_{|\varepsilon_n|<\Omega_D} (u_{n,\downarrow}^{\bm \eta} (\bm p)v_{n,\uparrow}^{\bm \eta*}(\bm p)(1-f_n)+u_{n,\uparrow}^{\bm \eta}(\bm p) v_{n,\downarrow}^{\bm \eta*}(\bm p)f_n),
\label{self-consistency_BdG}
\end{align}
\begin{align}
F_{\bm \eta,\alpha\beta}(\omega_m)= \int\limits_{BZ} \frac{d(pa)^{D-1}}{\pi^{D-1}}\sum\limits_n \left(\frac{  u_{n,\alpha}^{\bm \eta}(\bm p) v_{n,\beta}^{\bm \eta*}(\bm p)}{i \omega_m -\varepsilon_n(\bm p)}+\frac{ u_{n,\beta}^{\bm \eta}(\bm p) v_{n,\alpha}^{\bm \eta*}(\bm p)}{i \omega_m +\varepsilon_n(\bm p)} \right) .
\end{align}

\subsubsection{Numerical results for triplet correlations in a 2D S/AF heterostructure}
\label{numerical_2D}

\begin{figure}[tbh!]
\centering
\includegraphics[width=0.6\columnwidth]{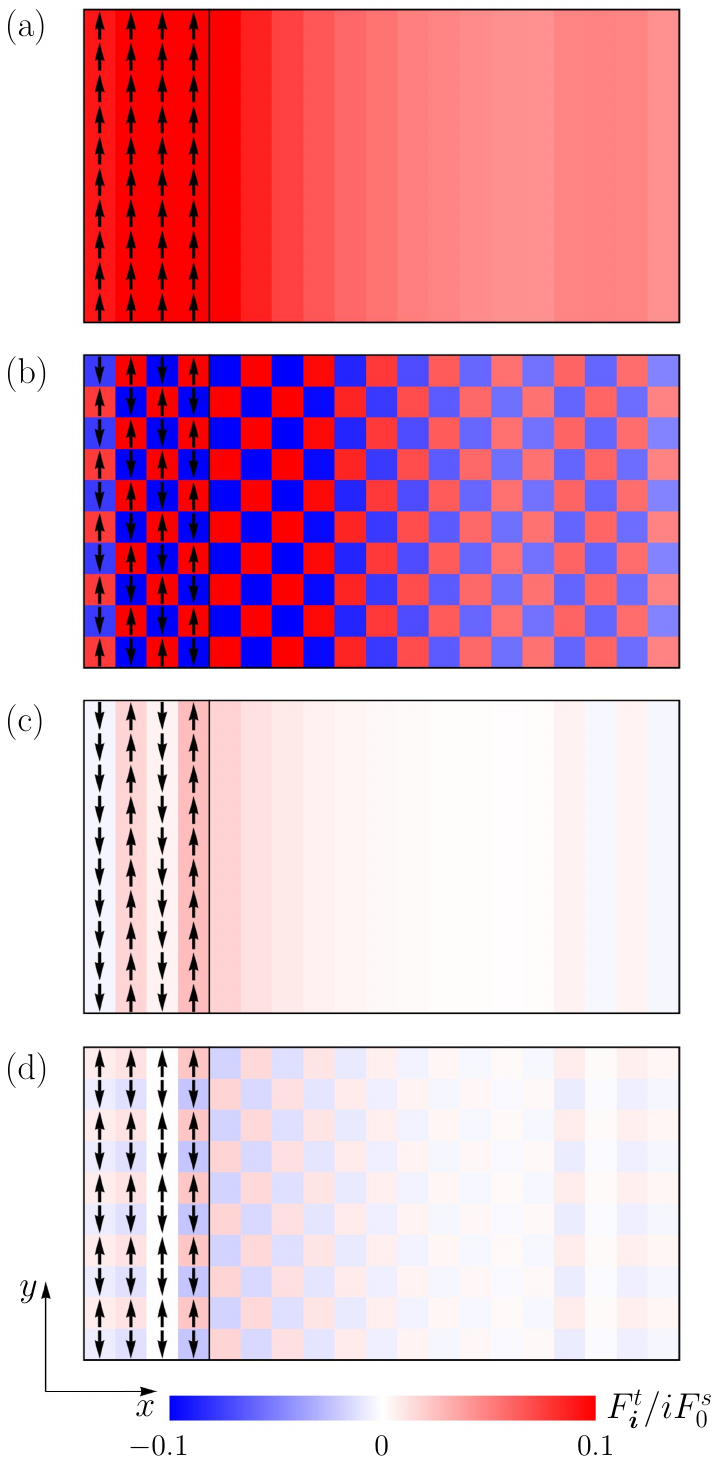}
\caption{Triplet correlations at 2D magnet/superconductor interfaces with different types of magnetic order in the magnetic material. Each cell represents one site. The system is infinite along the $y$-axis. The full number of sites along the $x$-axis is $N_{AF}=4$ for the magnetic part and $N_S=15$ for the superconducting part, as it is shown in the figures. The amplitude of the triplet correlations calculated according to Eq.~(\ref{triplet_total}) is shown by color. The magnet occupies 4 left rows of the system, the type of magnetic ordering is shown by arrows. $F^s_0$ is the amplitude of singlet correlations, calculated according to Eq.~(\ref{triplet_total}) for the isolated superconductor. The superconducting OP of the isolated superconductor $\Delta_0=0.035t$, $h=0.04t$, $\Omega_D=0.5t$, $T=0.005t$, $\mu=0$.} 
 \label{fig:2D}
\end{figure}

The triplet correlations induced at the superconductor/magnet interface having different types of magnetic ordering in the 2D case are demonstrated in Fig.~\ref{fig:2D}. It is seen that the S/F interface [Fig.~\ref{fig:2D}(a)] produce conventional triplet correlations slightly decaying into the depth of the superconductor. As it is expected, the decay length is of the order of $\xi_S$, which in our case is $\xi_S = v_F/2 \pi T_c = 16a$. The decay is superimposed by weak oscillations. The reason for the oscillations is the interference of the correlations at the momentum trajectories incoming to the impenetrable S layer edge and outgoing from it. The  S/AF interface with checkerboard magnetic order in the AF generates the  N\'eel-type checkerboard triplet correlations [Fig.~\ref{fig:2D}(b)] with approximately the same intensity \cite{Bobkov2022} as for the case of the S/F interface and the same decay length superimposed by weak oscillations of the same physical nature. 

Figs.~\ref{fig:2D}(c) and (d) represent the triplet correlations induced at the S/AF interfaces with compensated layered antiferromagnet having two different orientations with respect to the interface. If the ferromagnetic layers are oriented along the interface [Fig.~\ref{fig:2D}(c)], the proximity induced triplet correlations are {\it conventional triplet sign-preserving correlations}. The only difference from the case of S/F interface is that their amplitude is strongly weakened due to the fact that the correlations, which are produced by subsequent layers of the antiferromagnet, have the opposite sign and thus partially cancel each other. The cancellation is not full because the ferromagnetic layers, which are far from the interface, produce weaker triplet correlations, as it should be expected. The most unexpected thing happens when we consider S/AF interface with ferromagnetic layers oriented perpendicular to the interface [Fig.~\ref{fig:2D}(d)]. In this case each of the AF layers aligned with the interface produce {\it the N\'eel-type checkerboard} triplet correlations. Analogously to Fig.~\ref{fig:2D}(c) the correlations produced by subsequent layers also partially cancel each other resulting in much weaker amplitude of the N\'eel triplet correlations. 

Thus, the results obtained for 2D S/AF heterostructures are in agreement with the results for homogeneous antiferromagnetic superconductors, discussed in Sec.~\ref{homogeneous}: only checkerboard N\'eel triplet correlations are visible, and layered sign-changing triplet correlations are not observed. More precisely, the layered correlations are produced, but they are small in parameter $h/t$ and for this reason are masked by much stronger checkerboard N\'eel triplet correlations. Please note, that for homogeneous antiferromagnetic superconductors the type of proximity-generated N\'eel triplet correlations must coincide with the type of the underlying AF order. For this reason the layered antiferromagnets do not produce noticeable N\'eel triplet correlations. However, in spatially separated S/AF heterostructures this symmetry restriction is removed and S/AF interfaces with layered AFs can also produce conventional triplet correlations or checkerboard  N\'eel triplet correlations, the amplitude of which does not contain a small multiplier $h/t$, provided that the nesting is good. For this reason we observe these types of correlations at S/AF interfaces with layered antiferromagnets. 

With increasing the on-site energy $\mu$ at the same value of $h$ the N\'eel triplet correlations get weaker due to the destroying the nesting condition, and the amplitude of the conventional triplet correlations remains practically the same. The presence of a tunnel barrier at the interface weakens the magnitude of all triplet correlations equally. 

\subsubsection{Numerical results for triplet correlations in a 3D S/AF heterostructure}
\label{numerical_3D}

\begin{figure}[tbh!]
\centering
\includegraphics[width=0.6\columnwidth]{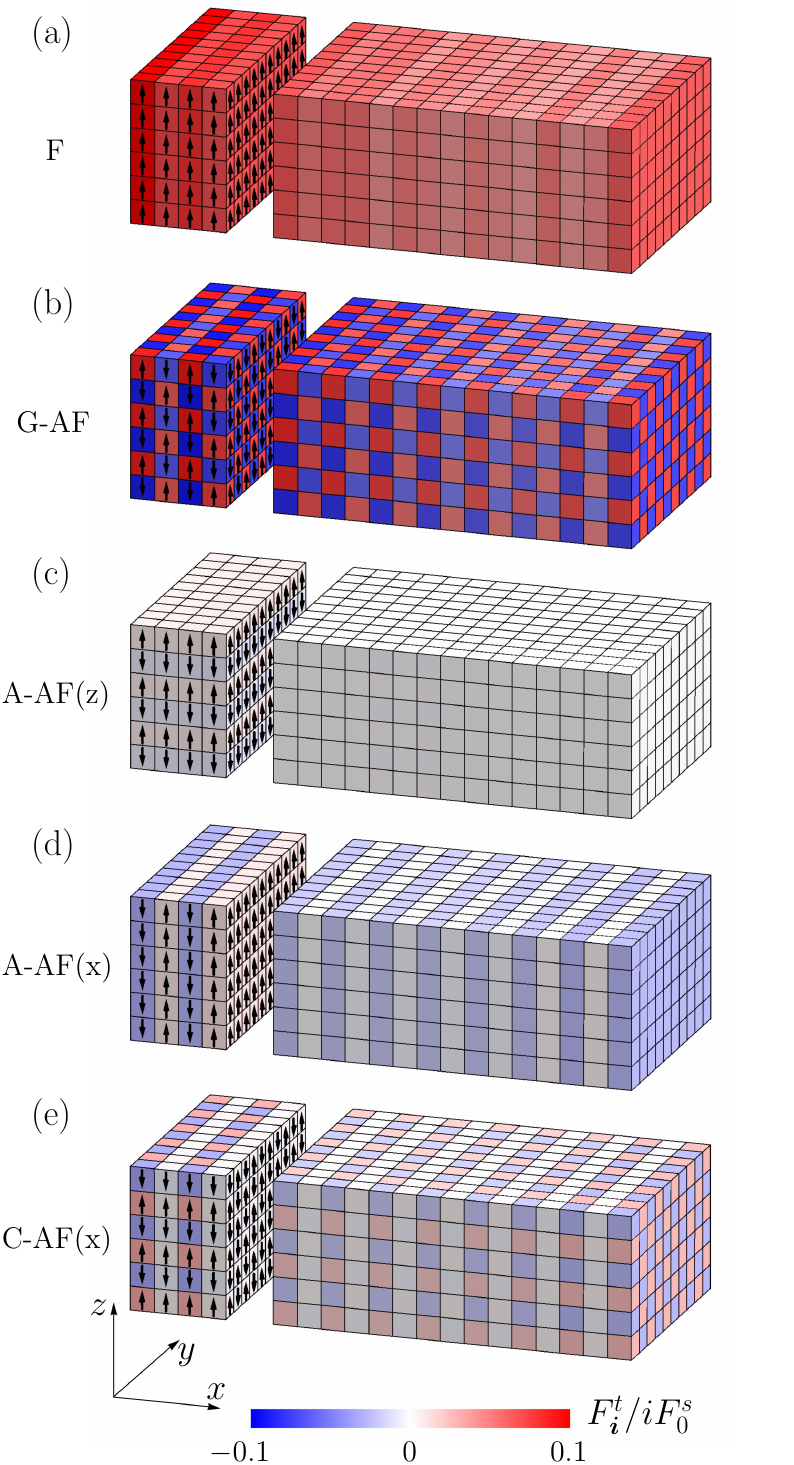}
\caption{Triplet correlations at 3D magnet/superconductor interfaces with different types of magnetic order in the magnetic material. Each cell represents one site. The system is infinite in the $(y,z)$-plane. The real size of the system along the $x$-axis is shown in the figures. The amplitude of the triplet correlations calculated according to Eq.~(\ref{triplet_total}) is shown by color. The magnet occupies 4 left planes of the system, the type of magnetic ordering is shown by arrows.  $h=0.05t$, other parameters are the same as Fig. \ref{fig:2D}.} 
 \label{fig:3D}
\end{figure}

The triplet correlations induced at the superconductor/magnet interface having different types of magnetic ordering in the 3D case are demonstrated in Fig.~\ref{fig:3D}. In the 3D case we consider $A$-type and $C$-type AF orders. Similar to the 2D case the S/F interface [Fig.~\ref{fig:3D}(a)] produce conventional triplet correlations slightly decaying into the depth of the superconductor at $\xi_S$, which is again superimposed by the interference pattern of the momentum trajectories incoming to the impenetrable S layer edge and outgoing from it. 3D  S/AF interface with $G$-type magnetic order generates the  N\'eel-type checkerboard triplet correlations [Fig.~\ref{fig:2D}(b)] decaying at $\xi_S$ with approximately the same intensity as for the case of the S/F interface. 

Figs.~\ref{fig:3D}(c) and \ref{fig:3D}(d) demonstrate triplet correlations obtained at 3D S/AF interfaces with layered $A$-type AFs in two different orientations with respect to the S/AF interface. In Fig.~\ref{fig:3D}(c) the normal to the plane of ferromagnetic layers is along the S/AF interface. It is seen that in contrast to the 2D S/AF interface with layers oriented perpendicular to the interface [Fig.~\ref{fig:2D}(d)] the triplet correlations are not visible at all. This is because the layered AF order in plane of the interface prevents generation of the checkerboard N\'eel triplet correlations, and the layered $A$-type N\'eel triplet correlations are negligibly small, as it was discussed above. It is interesting to compare this situation with the case of S/AF interface with $C$-type antiferromagnet oriented such that the normal to the antiferromagnetic layers is perpendicular to the S/AF interface (along the $x$-axis), see Fig.~\ref{fig:3D}(e). Both situations, presented in Figs.~\ref{fig:3D}(c) and (e) correspond to the same magnetic ordering in the $(x,z)$-plane, but differ by the magnetic ordering in plane of the S/AF interface, which is layered for Fig.~\ref{fig:3D}(c) and checkerboard antiferromagnetic for Fig.~\ref{fig:3D}(e). In the second case we see that the checkerboard N\'eel triplet correlations are induced because the checkerboard AF ordering of the interface allows for their generation. However, the overall amplitude of these checkerboard N\'eel triplet correlations is weaker than for the S/AF interface with $G$-type antiferromagnet presented in Fig.~\ref{fig:3D}(b) in full analogy with the 2D case considered before. 

In Fig.~\ref{fig:3D}(d) the normal to the plane of ferromagnetic layers of the $A$-type antiferromagnet is perpendicular to the S/AF interface. In full analogy with the 2D case [see Fig.~\ref{fig:2D}(c)] weak conventional sign preserving triplet correlations are produced here. The periodic spatial structure of their amplitude is due to the inevitable Friedel-like oscillations of the anomalous Green's function. The same Friedel-like oscillations are also seen on top of the checkerboard N\'eel triplet superconducting correlations in Fig.~\ref{fig:2D}(e). For other orientations of the $C$-type AF with respect to the interface the situation should be similar to the presented in Fig.~\ref{fig:3D}(c), that is noticeable triplet correlations are not produced because the layered (not checkerboard) antiferromagnetic ordering in plane of the interface prevents their generation.

\section{Superconducting order parameter in S/AF heterostructures: comparison between different types of AF ordering}
\label{OP}

\begin{figure}[tbh!]
\centering
\includegraphics[width=0.8\columnwidth]{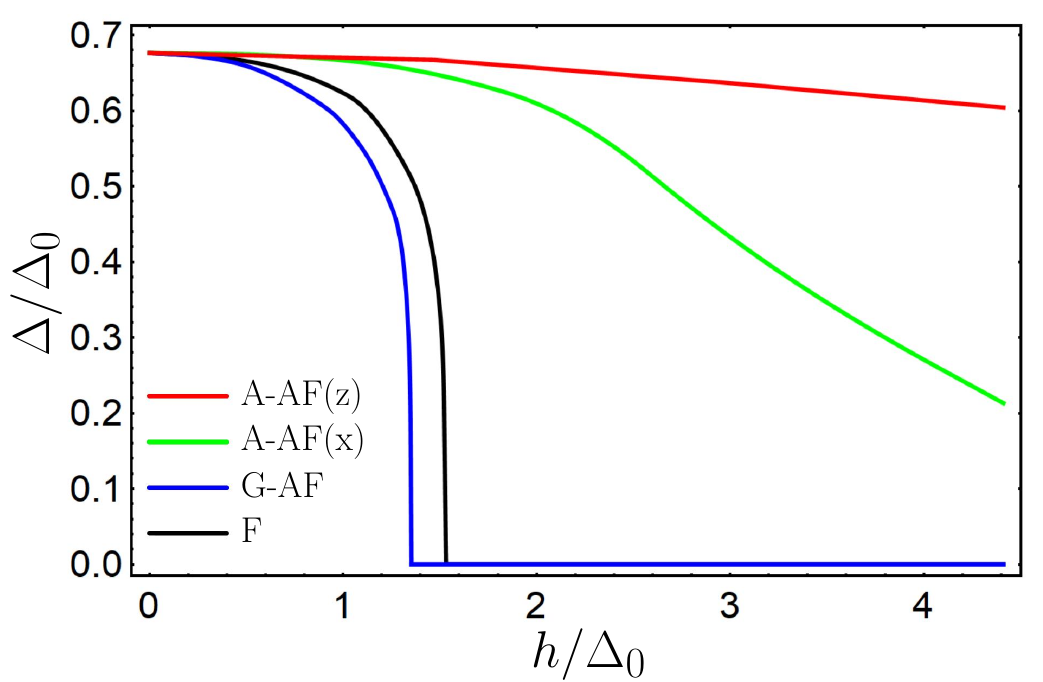}
\caption{Superconducting order parameter as a function of the exchange field of the magnet. The value of the OP is averaged over all superconducting sites. Different curves correspond to different 3D superconductor/magnet heterostructures presented in Fig.~\ref{fig:3D}. Black - S/F heterostructure [Fig.~\ref{fig:3D}(a)]; blue - S/AF heterostructure with $G$-type AF [Fig.~\ref{fig:3D}(b)]; green -  S/AF heterostructure with $A$-type AF with normal to the layers oriented along the $x$-axis [Fig.~\ref{fig:3D}(d)]; red - S/AF heterostructure with $A$-type AF with normal to the layers oriented along the $z$-axis [Fig.~\ref{fig:3D}(c)]. $\Delta_0=0.035t$, $\Omega_D=0.5t$, $T=0.005t$, $\mu=0$.} 
 \label{fig:OP}
\end{figure}

In this section we discuss the influence of the discussed proximity effect on the superconducting order parameter in 3D S/AF heterostructures. The superconducting OP was calculated self-consistently according to Eq.~(\ref{self-consistency_BdG}) for the systems sketched in Fig.~\ref{fig:3D}. Because of the singlet-triplet conversion the exchange field of the magnet should suppress the superconducting OP. Since the length of the superconducting layer is shorter than $\xi_S$, the spatial variation of the OP is weak. For this reason we present the OP averaged over all superconducting sites. It is presented in Fig.~\ref{fig:OP} as a function of the exchange field of the magnet. The suppression is the most pronounced for S/F systems and for S/AF systems with $G$-type antiferromagnets. It is in agreement with the fact that the proximity-induced triplet correlations are the strongest for these types of magnetic orderings, and both types of triplet correlations, conventional and checkerboard N\'eel-type, suppress the superconductivity equally. The suppression is a bit stronger for the S/AF interface with $G$-type AF than for the S/F interface, what was already reported in the literature \cite{Bobkov2022} and is explained by the fact that at half-filling there is a gap in the electronic density of states of the antiferromagnet exactly at the Fermi level, what results in the reduced number of electronic states available for superconducting pairing.

The suppression of the OP in S/AF heterostructures with $A$-type antiferromagnet is weaker, as it is expected from the consideration of triplet correlations induced in such systems. If the normal to the plane of ferromagnetic layers is oriented perpendicular to the interface (green curve in Fig.~\ref{fig:OP}), then according to Fig.~\ref{fig:3D}(d) weak conventional triplet correlations are induced. They suppress superconductivity, but the degree of the suppression is much weaker than for S/F heterostructures because of the lower amplitude of triplets in this case. If the normal to the plane of ferromagnetic layers is oriented parallel to the interface (red curve in Fig.~\ref{fig:OP}), then according to Fig.~\ref{fig:3D}(c) there no noticeable triplets in the system. Consequently, we observe very weak suppression of the superconducting OP.

It is worth mentioning that for all curves presented in Fig.~\ref{fig:OP} the superconducitng OP at zero exchange field of the AF is smaller than the OP of the isolated superconductor, that is $\Delta(h=0)<\Delta_0$. The reason is that at $h=0$ the system is not equivalent to an isolated superconducting film with $\Delta = \Delta_0$. It is rather equivalent to the superconductor/normal metal  heterostructure. The observed suppression of $\Delta$ with respect to $\Delta_0$ at $h=0$ is due to the leakage of the superconducting correlations into the N region.

Although not presented in the figures, the suppression of the OP in homogeneous antiferromagnetic superconductors manifest the same general trends. The strongest suppression occurs in ferromagnetic superconductors \cite{Sarma1963} and in antiferromagnetic superconductors with $G$-type magnetic ordering at half-filling \cite{Bobkov2022}. However, in antiferromagnetic superconductors with layered magnetic ordering the suppression is very weak and the superconductivity can be fully destroyed only by extremely high exchange fields of the order of the Fermi energy, as it was already reported in \cite{Sukhachov2025}. The only difference of the antiferromagnetic superconductors from the thin-film S/AF heterostructures is the absence of the leakage of the superconducting correlations into the nonsuperconducting region.

\section{Conclusions}
\label{conclusions}

In this paper we investigated the types and magnitude of the N\'eel triplet superconducting correlations, which are induced in homogeneous antiferromagnetic superconductors and S/AF heterostructures with an arbitrary two-sublattice antiferromagnet. The superconductivity suppression resulting from the singlet-triplet conversion is also studied. It is shown that the appearance of the N\'eel triplet correlations is very sensitive to the
type of the antiferromagnetic order. In antiferromagnetic superconductors and S/AF heterostructures with $G$-type antiferromagnetic order near half-filling the checkerboard N\'eel triplet correlations with approximately the same amplitude as conventional triplets in S/F heterostructures are induced. At the same time, if the exchange field of the antiferromagnet is small as compared to the conduction band width,  $A$-type or $C$-type N\'eel triplet correlations of noticeable amplitude are never generated in antiferromagnetic
superconductors and S/AF heterostructures with layered antiferromagnets of $A$ and $C$-type. The amplitude of such correlations is proportional to the small factor $h/t$. The general reason of such discrepancy is the impossibility to reach a good nesting condition in S/AF heterostructures with antiferromagnets of such types at any parameters of the materials. The strong weakening of the $A$-type or $C$-type N\'eel triplet correlations at reasonable values of the exchange field with respect to the checkerboard N\'eel triplet correlations leads to the fact that the conventional singlet superconductivity is suppressed by the layered antiferromagnetic order to a much lesser extent than by the checkerboard  antiferromagnetic order at the same value of the exchange field. 

Thin-film S/F and S/AF heterostructures (the term "thin-film" means that the superconducting layer thickness does not exceed the superconducting coherence length $\xi_S$) are frequently described in the framework of an effective model, when the heterostructure is changed by a homogeneous superconductor with the exchange field of the specific type. In the present paper we demonstrated that in S/AF heterostructures with layered antiferromagnets only conventional triplet correlations or checkerboard N\'eel correlations are induced depending on the orientation of the antiferromagnet with respect to the S/AF interface. This means that one should be careful when choosing an effective homogeneous model describing a heterostructure. For example, as follows from the above, a S/AF heterostructure with a layered antiferromagnet within the framework of an effective model can correspond to a ferromagnetic order or a checkerboard antiferromagnetic order of the effective exchange field of a homogeneous antiferromagnetic superconductor. 

Since the triplet correlations is the cornerstone of the majority of effects of superconducting spintronics, the presented study of the superconducting proximity effect and the N\'eel triplet correlations in S/AF heterostructures is of fundamental importance for an efficient integration of antiferromagnetic materials into the field of superconducting spintronics. In addition, our results indicate that it is better to choose layered rather than checkerboard antiferromagnets for heterostructures if the appearance of triplet correlations is unimportant for the effects under consideration and suppression of superconductivity is undesirable. This situation may arise if the antiferromagnet is used, for example, only to provide an exchange bias for a ferromagnet in some types of planar heterostructures. Another relevant case is if only the electromagnetic interaction between the superconductor and the antiferromagnet is important. Furthermore, our results suggest that the coexistence of antiferromagnetism and $s$-wave superconductivity should rather be characteristic of layered, and not checkerboard antiferromagnets.

\bmhead{Acknowledgements}

The financial support from the Russian Science
Foundation via the project No.24-22-00186 is acknowledged. 

\section*{Declarations}

{\bf Competing interests.} The authors declare no competing interests.

\bibliography{striped}

\end{document}